\documentclass[aps,prl,twocolumn,superscriptaddress,showpacs,letterpaper,nofootinbib,floatfix]{revtex4}

\usepackage{graphicx}

\begin{document}


\title{The Beam--Charge Azimuthal Asymmetry and Deeply Virtual Compton Scattering}


\def\groupalberta{\affiliation{Department of Physics, University of Alberta, Edmonton, Alberta T6G 2J1, Canada}}
\def\groupargonne{\affiliation{Physics Division, Argonne National Laboratory, Argonne, Illinois 60439-4843, USA}}
\def\groupbari{\affiliation{Istituto Nazionale di Fisica Nucleare, Sezione di Bari, 70124 Bari, Italy}}
\def\groupbeijing{\affiliation{School of Physics, Peking University, Beijing 100871, China}}
\def\groupchina{\affiliation{Department of Modern Physics, University of Science and Technology of China, Hefei, Anhui 230026, China}}
\def\groupcolorado{\affiliation{Nuclear Physics Laboratory, University of Colorado, Boulder, Colorado 80309-0390, USA}}
\def\groupdesy{\affiliation{DESY, 22603 Hamburg, Germany}}
\def\groupzeuthen{\affiliation{DESY, 15738 Zeuthen, Germany}}
\def\groupdubna{\affiliation{Joint Institute for Nuclear Research, 141980 Dubna, Russia}}
\def\grouperlangen{\affiliation{Physikalisches Institut, Universit\"at Erlangen-N\"urnberg, 91058 Erlangen, Germany}}
\def\groupferrara{\affiliation{Istituto Nazionale di Fisica Nucleare, Sezione di Ferrara and Dipartimento di Fisica, Universit\`a di Ferrara, 44100 Ferrara, Italy}}
\def\groupfrascati{\affiliation{Istituto Nazionale di Fisica Nucleare, Laboratori Nazionali di Frascati, 00044 Frascati, Italy}}
\def\groupgent{\affiliation{Department of Subatomic and Radiation Physics, University of Gent, 9000 Gent, Belgium}}
\def\groupgiessen{\affiliation{Physikalisches Institut, Universit\"at Gie{\ss}en, 35392 Gie{\ss}en, Germany}}
\def\groupglasgow{\affiliation{Department of Physics and Astronomy, University of Glasgow, Glasgow G12 8QQ, United Kingdom}}
\def\groupillinois{\affiliation{Department of Physics, University of Illinois, Urbana, Illinois 61801-3080, USA}}
\def\groupmichigan{\affiliation{Randall Laboratory of Physics, University of Michigan, Ann Arbor, Michigan 48109-1040, USA }}
\def\groupmoscow{\affiliation{Lebedev Physical Institute, 117924 Moscow, Russia}}
\def\groupnikhef{\affiliation{Nationaal Instituut voor Kernfysica en Hoge-Energiefysica (NIKHEF), 1009 DB Amsterdam, The Netherlands}}
\def\groupstpetersburg{\affiliation{Petersburg Nuclear Physics Institute, St. Petersburg, Gatchina, 188350 Russia}}
\def\groupprotvino{\affiliation{Institute for High Energy Physics, Protvino, Moscow region, 142281 Russia}}
\def\groupregensburg{\affiliation{Institut f\"ur Theoretische Physik, Universit\"at Regensburg, 93040 Regensburg, Germany}}
\def\grouprome{\affiliation{Istituto Nazionale di Fisica Nucleare, Sezione Roma 1, Gruppo Sanit\`a and Physics Laboratory, Istituto Superiore di Sanit\`a, 00161 Roma, Italy}}
\def\groupsimonfraser{\affiliation{Department of Physics, Simon Fraser University, Burnaby, British Columbia V5A 1S6, Canada}}
\def\grouptriumf{\affiliation{TRIUMF, Vancouver, British Columbia V6T 2A3, Canada}}
\def\grouptokyo{\affiliation{Department of Physics, Tokyo Institute of Technology, Tokyo 152, Japan}}
\def\groupamsterdam{\affiliation{Department of Physics and Astronomy, Vrije Universiteit, 1081 HV Amsterdam, The Netherlands}}
\def\groupwarsaw{\affiliation{Andrzej Soltan Institute for Nuclear Studies, 00-689 Warsaw, Poland}}
\def\groupyerevan{\affiliation{Yerevan Physics Institute, 375036 Yerevan, Armenia}}
\def\groupnone{\noaffiliation}


\groupalberta
\groupargonne
\groupbari
\groupbeijing
\groupchina
\groupcolorado
\groupdesy
\groupzeuthen
\groupdubna
\grouperlangen
\groupferrara
\groupfrascati
\groupgent
\groupgiessen
\groupglasgow
\groupillinois
\groupmichigan
\groupmoscow
\groupnikhef
\groupstpetersburg
\groupprotvino
\groupregensburg
\grouprome
\groupsimonfraser
\grouptriumf
\grouptokyo
\groupamsterdam
\groupwarsaw
\groupyerevan


\author{A.~Airapetian}  \groupmichigan
\author{N.~Akopov}  \groupyerevan
\author{Z.~Akopov}  \groupyerevan
\author{M.~Amarian}  \groupzeuthen \groupyerevan
\author{A.~Andrus}  \groupillinois
\author{E.C.~Aschenauer}  \groupzeuthen
\author{W.~Augustyniak}  \groupwarsaw
\author{R.~Avakian}  \groupyerevan
\author{A.~Avetissian}  \groupyerevan
\author{E.~Avetissian}  \groupfrascati
% ...

\collaboration{The HERMES Collaboration} \noaffiliation




\date{\today}

\begin{abstract}
The first observation of an azimuthal cross--section asymmetry 
with respect to the charge of the
incoming lepton beam is reported 
from a study of hard exclusive electroproduction of real photons.
The data have been accumulated by the HERMES experiment at DESY,
in which the HERA 27.6 GeV electron or positron beam scattered
off an unpolarized hydrogen gas target.
The observed asymmetry is attributed to the interference between 
the Bethe--Heitler process and the
Deeply Virtual Compton Scattering (DVCS) process.
The interference term is sensitive to 
DVCS amplitudes, which provide
the most direct access to 
Generalized Parton Distributions. 
\end{abstract}


\maketitle


The partonic structure of the nucleon has been traditionally described
in terms of Parton Distribution Functions (PDFs), which appear in the
interpretation of, e.g., inclusive Deeply Inelastic Scattering (DIS).  More
recently, PDFs have been subsumed within Generalized Parton Distributions
(GPDs)~\cite{Mue94, Ji97, Rad97}, which relate directly to hard exclusive processes that
involve at least one additional hard vertex, yet leave the target nucleon intact.
The ordinary PDFs and nucleon elastic form factors appear as
kinematic limits and moments of GPDs, respectively.
Strong interest in the formalism of GPDs
has emerged after GPDs were found to offer the first
possibility to reveal
the total angular momentum carried by the quarks in the nucleon~\cite{Ji97}.  
More recent discussions focus on 
the potential of GPDs as a three--dimensional representation of hadrons at the parton level 
\cite {Bur00,Die02,Ral02,Bel02b,Bur03}.

Among all practical probes, the DVCS process, 
i.e., the hard exclusive leptoproduction of a real photon \mbox{($e\, p \rightarrow
e\, p \, \gamma$)},
appears to provide the theoretically cleanest access to GPDs.
Direct access to the DVCS {\it amplitudes} is provided by 
the interference between the DVCS and Bethe--Heitler (BH) processes, 
in which the photon is radiated from a parton 
and from the lepton, 
respectively.
Since these processes have an identical final state, 
the squared photon production amplitude is given by
\begin{equation} \label {eqn:tau}
\left| \tau \right|^2 
= \left| \tau_{{\scriptscriptstyle BH}} \right|^2 + 
\left| \tau_{{\scriptscriptstyle DVCS}} \right|^2 + \underbrace{
\tau_{{\scriptscriptstyle DVCS}} \, \tau_{{\scriptscriptstyle BH}}^* 
+ \tau_{{\scriptscriptstyle DVCS}}^* \, \tau_{{\scriptscriptstyle BH}}}_I,
\end{equation}
where $I$ denotes the interference term.
It introduces a dependence on the beam charge, which is a rare phenomenon
normally confined to processes involving the weak interaction.
The BH amplitude ($\tau_{{\scriptscriptstyle BH}}$) is precisely
calculable from measured elastic form factors. 
The cross section 
depends on the Bjorken scaling variable $x_B$, 
the squared virtual--photon four--momentum $-Q^2$,
and the squared four-momentum 
transfer $t$ to the target.
This interpretation of the virtual--photon kinematics does not apply to the BH process.
In addition, the cross section depends on the azimuthal angle $\phi \in
[-\pi,\pi]$, defined as the angle between the plane containing the incoming and outgoing lepton trajectories
and the plane correspondingly defined by the virtual and real photon~\cite{Bac04}.

For an unpolarized proton target, and $-t << Q^2$, the interference term is given by \cite{Die97}
\begin{eqnarray} \label {I}
I \propto 
-C \, 
[ \, a \, \cos \phi \, \mathrm{Re} \widetilde {\cal M}^{1,1}
- b \, P_l \, \sin \phi  \, \mathrm{Im} \widetilde {\cal M}^{1,1} 
\, ], 
\end{eqnarray}  
where 
the lepton beam has longitudinal polarization $P_l$
and charge $C = \pm 1$,
and $a$ and $b$ are functions of the ratio of longitudinal to transverse 
virtual--photon flux.
A polarization--independent constant term, higher harmonics 
($\cos 2\phi$, $\cos 3\phi$, $\sin 2\phi$), as well as 
a $\cos \phi$ dependence in the prefactor,
have been neglected since they 
are suppressed by at least $\mathcal{O}(1/Q)$ or $\mathcal{O}(\alpha_s)$. 
Here $\alpha_s$ is the strong coupling constant. 
The squared BH and DVCS amplitudes have their own
$\phi$ dependences, but do not depend on the sign of the charge.
Hence the measurement of a cross section asymmetry with respect to the beam
charge
$(d\sigma^+ -d\sigma^-)/ (d\sigma^+ + d\sigma^-)$
is a way to single out the interference term \cite {Bro72} in the numerator, 
while the denominator is dominated by a $\phi$--independent BH contribution. 
See  Ref.~\cite {Bel02a} for details and full equations.

The photon--helicity--conserving amplitude  
\begin{equation} \label{m11}
\widetilde {\cal M}^{1,1}=F_1 \, {\cal H} + \frac{x_B}{2-x_B}(F_1 
+F_2) \, \widetilde {\cal H} - \frac{t}{4 M_p^2} F_2 \, {\cal E}
\end{equation}
is given by a linear combination of the Compton Form Factors (CFFs)
${\cal H}$, $\widetilde {\cal H}$ and ${\cal E}$, together with
the Dirac and Pauli form factors $F_1$ and $F_2$~\cite {Bel02a}.
Here $M_p$ denotes the proton mass.
The CFFs are convolutions of the corresponding twist--2 GPDs $H$, $\widetilde H$ and $E$
with the hard scattering amplitude.

The $\sin \phi$ modulation accessing the imaginary part 
of $\widetilde {\cal M}^{1,1}$ has already been observed
in cross--section asymmetries with respect to the beam 
helicity~\cite {Air01, Ste01}.
This Letter reports the first measurement
of an asymmetry with respect to the beam 
charge, accessing the real part
of $\widetilde {\cal M}^{1,1}$ via a $\cos \phi$ modulation.

Data with an unpolarized hydrogen target were 
accumulated using 
the HERMES spectrometer \cite {Ack98} 
and the longitudinally 
polarized 27.6 GeV electron and positron beams
of the HERA accelerator at DESY.
Events were selected if they contained exactly one photon and one charged track 
identified as the scattered lepton. 
The hadron contamination in this lepton sample was kept below 1\% by combining
the information from a 
transition--radiation detector, a preshower scintillator detector, and 
an electromagnetic calorimeter.
The kinematic requirements imposed 
were 1~GeV$^2 < Q^2 <$ 10~GeV$^2$, 0.03 $< x_B <$ 0.35, 
$W >$ 3 GeV, and $\nu <$ 22 GeV, where $W$ denotes the initial photon--nucleon 
invariant mass and $\nu$ is the virtual--photon energy in the target rest frame.
The real photon was identified by detecting an energy deposition above 5~GeV in 
the calorimeter in addition to a signal in the preshower detector, 
without an associated charged track in the back region of the spectrometer.
Unlike in a preliminary stage of this analysis \cite {Ell02b}, 
the useful range of the polar angle 
$\theta_{\gamma^* \gamma}$ between the virtual and real photons is 
not limited by the photon position resolution,
due to an improved reconstruction algorithm.
Now the 
restriction on $\theta_{\gamma^* \gamma}$ is relaxed to $\theta_{\gamma^* \gamma} > 5$~mrad,
limited mainly by the electron momentum resolution.
In addition, a stricter upper limit of $\theta_{\gamma^* \gamma} < 45$~mrad 
is imposed in order to improve the 
signal--to--background ratio~\cite {Ell04}.

The recoiling proton was not detected. 
Hence exclusive events are selected by requiring the 
missing mass $M_X$ of the reaction $e p \rightarrow e \gamma X$ 
to be close to the proton mass.
\begin{figure}[t!]
  \includegraphics[width=\columnwidth]{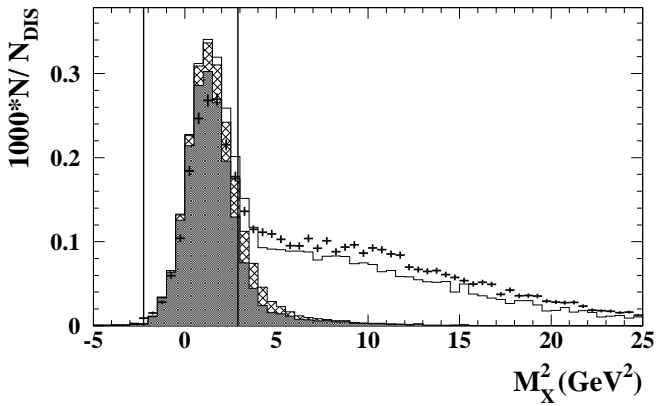}
   \caption{Distributions in missing--mass squared from 
data (statistical error bars) 
and from Monte Carlo simulations (line). The latter include elastic BH (filled
area) and associated
BH (hatched area) processes as well as semi--inclusive background.
The vertical lines enclose the selected exclusive region.}
  \label{missmass2}
\end{figure}
Fig.~\ref {missmass2} shows the distribution in
$M_X^2 = ( q + p - q^\prime)^2$,
with $q$, $p$, and $q^\prime$ being the four--momenta of the virtual photon, 
the target nucleon in the initial state, and the real photon, respectively.
Mainly due to the resolution in photon energy, the exclusive peak 
extends to negative values, in which case
$M_X$ is defined as $-\sqrt{-M_X^2}$. 
The exclusive region is defined as $(-1.5$~GeV)$^2$ $< M_X^2 < (1.7$~GeV)$^2$, based on 
the result of a Monte Carlo simulation (MC), shown in the same figure. 
The Mo and Tsai formalism \cite {Mo69} is used to simulate the elastic BH
process leaving the target nucleon intact, and the associated BH process,
where the nucleon is excited to a resonant state.
For the latter a cross--section parametrization for 
the resonance region is used \cite {Bra76}. 
Not included in the simulation is the DVCS contribution, 
which in this kinematic regime is expected to be
much smaller than that of the BH process~\cite{Kor02a}.
The simulation also takes into account
the semi--inclusive production of neutral mesons (mostly $\pi^0$), 
where all but one of the decay photons escape detection. 
For this, the MC generator LEPTO \cite {Ing97} in conjunction with a special JETSET \cite {Sjo94} 
fragmentation tune is used, the latter being optimized for energies relevant to HERMES \cite {Hil05}.
Not shown in Fig.~\ref{missmass2} is the contribution of
exclusive $\pi^0$ production, which contributes less than 2.5\% to the
exclusive region 
based on the model in Ref.~\cite {Van99}.
As shown in Fig.~\ref {missmass2}, data and MC are in good agreement
taking into account that they are both absolutely normalized, and that
the MC does not include radiative corrections to the BH cross--section.
The MC yield exceeds the data by about 20\% in the exclusive region, as may be expected~\cite{Van00}.
A simulation including second order radiative processes should 
give an improved MC--data comparison in the full missing--mass 
region, since a part of the exclusive events experiencing second order radiative
corrections will not be lost but reconstructed in the non-exclusive region.

\begin{figure}[t!]
 \includegraphics[width=\columnwidth]{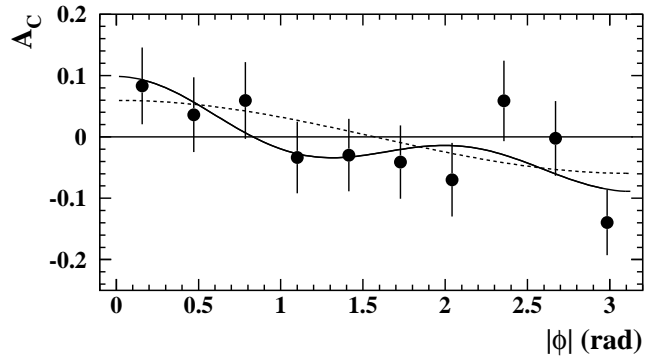}
  \caption{\label{bca_3p} Beam--charge asymmetry $A_C$ for the hard electroproduction
   of photons off protons as a function of the azimuthal angle $|\phi|$, for the
    exclusive sample
before background correction. Statistical uncertainties are shown.
The solid curve represents the four--parameter fit:
$(-0.011\pm0.019) + (0.060\pm0.027) \cos \phi + (0.016\pm0.026) 
\cos 2 \phi + (0.034\pm0.027) \cos 3 \phi$. 
The dashed line shows the pure $\cos \phi$ dependence.}
\end{figure}
\begin{figure}[t!]
 \includegraphics[width=\columnwidth]{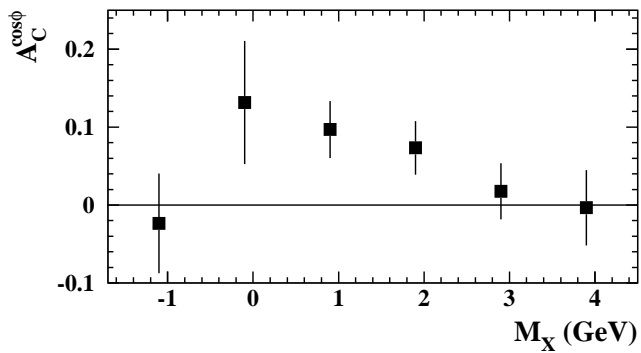}
  \caption{\label{cos_mx}
   The $\cos \phi$ amplitude of the beam--charge asymmetry as a 
  function of the missing mass, before background
  correction. Statistical uncertainties are shown.}
\end{figure}

The beam--charge asymmetry is evaluated as
\begin{equation} \label{xsec_bca_equation}
A_C(\phi) = \frac { N^+(\phi) - N^-(\phi)} {N^+(\phi) + N^-(\phi)},
\end{equation}
where $N^+(\phi)$ and $N^-(\phi)$ represent the single--photon yields per $\phi$ bin, 
normalized to the number of detected inclusive DIS events 
using the positron and electron beam, respectively.
Since these beams were polarized, sinusoidal contributions appear in numerator
and denominator of Eq.~\ref{xsec_bca_equation},
with the last term in Eq.~\ref{I} giving the biggest contribution.
In order to cancel these contributions, 
the `symmetrized' beam--charge asymmetry 
is calculated by replacing $\phi$ by $|\phi|$
in Eq.~\ref{xsec_bca_equation}.
The result for the exclusive region is displayed in Fig.~\ref{bca_3p}. 
The shown four--parameter fit
yields a non-zero $\cos \phi$ amplitude of $0.060 \pm 0.027$.
The result after the background correction described below 
is given in Table \ref{table:bca_c1}, last row.
The constant term as well as the $\cos 2 \phi$
and $\cos 3 \phi$ terms are compatible with zero.
Fig.~\ref {cos_mx} shows the $\cos \phi$ amplitude in 
several $M_X$ bins.
At higher $M_X$ the result is compatible with zero, confirming the absence of
charge--dependent instrumental effects.

As the recoiling proton remains undetected, $t$ is 
inferred from the 
measurement of the other final--state particles.
For elastic events, kinematics relate the energy with the direction of the
real photon, opening the possibility to omit the 
real--photon energy, which is the quantity subject to larger 
uncertainty.
Thus the value of 
$t$ in the exclusive region is calculated as
\begin{equation} \label{tc}
t = \frac{-Q^2 - 2 \, \nu \, (\nu - \sqrt{\nu^2 + Q^2} \, \cos\theta_{\gamma^* \gamma })}
{1 + \frac{1}{M_p} \, (\nu - \sqrt{\nu^2 + Q^2} \, cos\theta_{\gamma^* \gamma })}.
\end{equation}
The error caused by applying this expression to
inelastic events
($\approx$17\% in the exclusive region)
is accounted for in the MC simulation that is used to calculate 
the fractional contribution of background processes 
per kinematic bin in $-t$ (see Ref.~\cite {Ell04} for details).

\begin{figure}[t!]
 \includegraphics[width=\columnwidth]{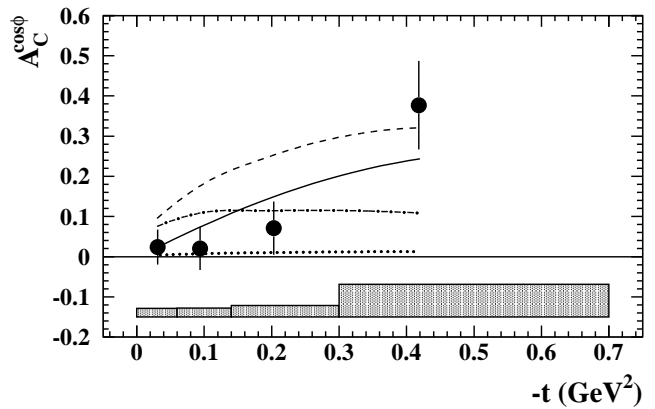}
   \caption{The $\cos \phi$ amplitude of the beam--charge asymmetry
    as a function of $-t$ for the exclusive region (-1.5~GeV $< M_X <$ 1.7~GeV), 
    after background correction.
    The error bars (band) represent(s) the statistical (systematic) uncertainties. 
    The calculations based on GPD models \cite {Van99,Goe01} use either a factorized 
    $t$--dependence with (dashed--dotted) or without (dotted)
    the D-term contribution, or a Regge--inspired $t$--dependence with (dashed) or without (solid)
    the D-term contribution.}
 \label{vgg_t_cos}
\end{figure}

Figure~\ref {vgg_t_cos} shows
the $\cos \phi$ amplitude derived from the four--parameter fit in each of four bins in $-t$.
In each bin, this result is corrected for the semi--inclusive background, which is treated
as a dilution since the background asymmetry can only be non-zero at
next-to-leading order in QED.
The total background contribution is about 6\%
as derived from the MC simulation, 
wherein the elastic and associated BH contributions are scaled down by 
the 20\% described above. 
The resulting $\cos \phi$ amplitudes are expected to originate from only 
elastic and associated production.
The associated BH processes contribute
about 5, 11, 18, and 29\% to the yields in the four $-t$ bins, or 
11\% in the full t--range,
with an estimated fractional uncertainty of 10\%.
The dominant contribution to the total
systematic uncertainty of the $\cos \phi$ amplitudes stems from effects 
due to possible deviations of the detector and/or the beam from their 
nominal positions. These effects can be as large as 0.02 per bin. 
Based on the models in Ref.~\cite{Kor02a},
acceptance and smearing effects can contribute up to 20\% of the $\cos \phi$
amplitude, and thus dominate the systematic uncertainty in the last $-t$ bin.
The other sources of systematic uncertainties are due to a
possible difference in
the calorimeter calibration between the two data sets, the
uncertainties from the semi-inclusive background correction
described above, and the dilution of the
asymmetry due to exclusively produced $\pi^0$ mesons misidentified as photons.
These contributions are combined quadratically in the total systematic uncertainty per 
bin in $-t$,
given in Table~\ref{table:bca_c1}\footnote{Note that a preliminary result
of this analysis~\cite {Ell02b} with a $t$--averaged value 
of $0.11 \, \pm \, 0.04$~(stat.) $\pm \, 0.03$~(sys.) was derived at a much
larger mean $-t$ value ($\langle \, -t \, \rangle$ = 0.27~GeV$^2$) due to different 
requirements on $\theta_{\gamma^* \gamma}$, as described above. 
It thus cannot be compared to the $t$--averaged result given in Table \ref{table:bca_c1} 
but approximately to the result in the third $-t$ bin.}.
Not included is any contribution due to additional QED vertices, as the most
significant of these has been estimated to be negligible, at least in the case
of polarization asymmetries~\cite{Afa05}.

The theoretical calculations for the $e \, p \rightarrow e \, p \, \gamma$
process shown 
in Fig.~\ref {vgg_t_cos} employ 
GPD models developed in 
Refs.~\cite {Van99,Goe01}, which are based on the
widely used framework of double distributions~\cite{Rad99}.
The model parameters of interest are those that change the GPD $H$
since the impact of the GPDs $\widetilde H$ and $E$ is suppressed at small
values of $x_B$ and $-t$, respectively (cf. Eq.~\ref{m11}).
The code of Ref.~\cite {Van01} was used to calculate the 
values for the 
$\cos \phi$ amplitude of the beam--charge asymmetry at the average kinematics
(see Table~\ref{table:bca_c1}) of every $-t$ bin and not at the kinematics of every event since
it is too computationally intensive.
\begin{table}
\begin{center}
\begin{tabular} {|c||c|c|c||r|} 
\hline
$-t$ bin & $ \langle \, -t \, \rangle $ & $ \langle \, x_B \, \rangle $ & $ \langle \, Q^2 \, \rangle $ & \rule {0mm} {4.5mm} $A_C^{\cos \phi} \pm$ stat. $\pm$ sys.~~~  \\
(GeV$^2$) & (GeV$^2$) &         & (GeV$^2$) & \\
\hline
\hline
~~~~~ $<$ 0.06 & 0.03 & 0.08 & 2.0 & 0.024 $\pm$ 0.043 $\pm$ 0.022   \\
\hline
0.06 -- 0.14 & 0.09 & 0.10 & 2.6 & 0.020 $\pm$ 0.054 $\pm$ 0.022   \\
\hline
0.14 -- 0.30 & 0.20 & 0.11 & 3.0 & 0.071 $\pm$ 0.066 $\pm$ 0.028   \\
\hline
0.30 -- 0.70 & 0.42 & 0.12 & 3.7 & 0.377 $\pm$ 0.110 $\pm$ 0.081   \\
\hline 
\hline 
~~~~~ $<$ 0.70 & 0.12 & 0.10 & 2.5 & 0.063 $\pm$ 0.029 $\pm$ 0.028   \\
\hline
\end{tabular}
\caption{The $\cos \phi$ amplitude of the beam--charge asymmetry per kinematic bin in
$-t$ after background correction and the respective average kinematic values.}
\label{table:bca_c1}
\end{center}
\end{table}
The difference between these two approaches is
strongly model dependent: Tests \cite {Kra05} show differences of up to 20\% using the models in
Ref.~\cite{Kor02a}, which are equivalent to the factorized models in Fig.~\ref
{vgg_t_cos}.
Four different parameter sets are selected by choosing
either a factorized or a Regge--inspired $t$--dependence, each with 
or without 
the contribution of a negative value of the so--called D-term~\cite{Pol99}, 
which is related to the spontaneous breaking of chiral symmetry in QCD~\cite{Goe01}. 
It contributes to the real part of the DVCS amplitude only and therefore can
be investigated with the beam--charge asymmetry for the first time.
A large positive D-term is ruled out by these data~\cite{Ell02b},
since it leads to a negative value for the beam--charge asymmetry.
The parameters $b_{val}$ and $b_{sea}$ in the profile 
function \cite {Mus00} are fixed at unity, since the beam-charge
asymmetry has been shown to be largely insensitive to them~\cite {Ell04}.
In comparing the predictions to the data at large $-t$, it should be borne in mind that
the model calculations do not include associated production, which
increases with $-t$ as mentioned above. 
The three data points at small $-t$ exclude the model based on the Regge--inspired
$t$-dependence with the D-term contribution.

These data are in agreement with a very recent calculation 
based on a dual parametrization of GPDs and either of two models for the
$t$-dependence~\cite{Guz06}.
Like the calculations shown in Fig.~\ref{vgg_t_cos}, these calculations
were performed at the average kinematics
(see Table~\ref{table:bca_c1}) of every $-t$ bin. 
Earlier model calculations were mostly done not as a function of $t$ but 
at the average kinematics of the preliminary result~\cite{Ell02b}, and 
thus cannot rigorously be compared to this measurement, see, 
e.g., Refs.~\cite{Kiv01a, Bel02a, Bel02b, Fre03}. However, since they span a wider or
different range for the magnitude of the beam--charge asymmetry 
when compared to the models shown here, it is apparent that this measurement 
can constrain the GPD~$H$.

In conclusion, a beam--charge azimuthal asymmetry in 
electroproduction of real
photons has been measured for the first time. 
A $\cos \phi$ dependence has been observed in a kinematic
region where the target proton is predominantly left intact. 
The $\cos \phi$ dependence is predicted to arise from the interference between the 
deeply virtual Compton scattering and Bethe--Heitler processes.  
These data can already be used to distinguish among theoretical models for
generalized parton distributions.

We gratefully acknowledge the DESY management for its support, the staff
at DESY and the collaborating institutions for their significant effort,
and our national funding agencies for financial support.

\bibliography{basename of .bib file}

\end{document}